# Synthesis and crystal growth of novel layered bismuthides $ATM_2Bi_2$, ($A$ = K, Rb; $TM$ = Zn, Cd) with significant electron-deficient structure.

Andrey I. Shilov [1,3], Kirill S. Pervakov [3], Konstantin A. Lyssenko [1], Vladimir A. Vlasenko [3], Dmitri V. Efremov [2], Bernd Büchner [2], Saicharan Aswartham [2], Igor V. Morozov [1]

[1] Lomonosov Moscow State University, Department of Chemistry, Moscow, Russia;
[2] Leibniz Institute for Solid State and Materials Research IFW Dresden, 01069 Dresden, Germany;
[3] P. N. Lebedev Physical Institute, Russian Academy of Sciences, 119991 Moscow, Russia;

## Abstract

Layered bismuthides of the family $ATM_2Bi_2$ are showing very interesting structural and physical properties. Here, we report crystal growth and characterization of new ternary layered bismuthides: $RbCd_2Bi_2$, $RbZn_2Bi_2$ and $KZn_2Bi_2$. Single crystals were grown by self-flux technique. As grown crystals were characterized by combination of powder X-ray diffraction, scanning electron microscopy, energy-dispersive X-ray spectroscopy and single crystal X-ray diffraction. All novel compounds crystallize in layered $ThCr_2Si_2$ structure type with plate-like morphology with weak bonding between layers. Significant structural difference between $RbCd_2Bi_2$ and $AZn_2Bi_2$ ($A$=K, Rb) was found.

**Keywords:** bismuthides, synthesis, crystal growth, crystal structure.

# 1. Introduction

Discovery of iron-based superconductors in ThCr$_2$Si$_2$ and PbFCl structure types [1] led the scientific community to search for new materials in the same families of compounds, such as 122 and 111 respectively. Despite of difference in composition, these structure types represent layered systems, where negatively charged antifluorite-like layers [*TPn*] (*T* - d-metal, *Pn*-pnictogen) with a significant contribution of the covalent bonding alternate with positively charged layers of cations (usually alkali or alkaline earth metals), which hold structure together by electrostatic forces. Compounds in these families are not bordered by superconductors, such as Ba(Fe, Co)$_2$As$_2$, RbFe$_2$As$_2$ or LiFeAs [2–5], they also contain thermoelectrics such as NaZnSb or CaZn$_2$Sb$_2$ [6,7] and non-trivial electronic properties as were found in BaMn$_2$Bi$_2$ and NaMnBi [8,9].

Compounds with 112 compositions have layered structure as well as 122 and 111 families but have one more layer in their structure consisting of pnictogen atoms linked in zigzag chains (*Pn* = As) or flat square networks (*Pn* = Bi). In addition to superconductors, this family is represented by compounds with non-trivial magnetic and electronic properties, like BaMnBi$_2$ [10] and Dirac semimetal candidates, as it was predicted in BaZnBi$_2$, but careful measurements showed its trivial nature in terms of exotic fermion states [11].

An exploratory materials search with bismuth as a pnictide in 111 and 122 families is an important step for new materials to search for the magnetic and non-trivial electronic properties. Indeed, due to high spin-orbit coupling (SOC) in bismuth, some of the layered compounds containing Bi can be unique in terms of their physical properties, as it was found giant rashba-splitting for BiTeI, topological insulator properties for Bi$_2$Te$_3$, 3D Dirac semimetal Na$_3$Bi [12–14]. Cadmium, which is 4d transition metal, can be considered as an alternative to replace Zn which is 3d-transition metal. The subtle balance between SOC and correlation can be interesting for the physical properties of the materials. As already known, Cd$_3$As$_2$ is an interesting Dirac semimetal with unusual electronic and transport properties.[15–17].

Recently, we have synthesized and investigated the structure of two new bismuthides, namely NaZnBi and NaCdBi. The first of which turned out to be isostructural to the 111 Fe-based superconductors family, and the second crystallizes in a different space group [18]. After some time, the compound NaZnBi was obtained in [19], and the authors concluded that there is a certain zinc deficiency in this structure. It is interesting to note that the synthesis of compounds Na*TM*Bi (T = Zn, Cd) confirmed the theoretical prediction of the existence of these compounds, but the theoretical calculations [20–22] predicted another crystal structure of these compounds.

Motivated by the theoretical prediction [21] and with our synthesis experience in *ATM*Bi we tried to synthesize *ATM*Bi, where *A* = K and Rb. However, in contrast to Na*TM*Bi, instead of 111 we obtained compounds with composition *ATM*$_2$Bi$_2$ which represent the 122 family. It should be noted that *AT*$_2$Bi$_2$ compounds in which A-alkaly metal remain practically unexplored: only a few examples of these compounds can be found in literature [23], despite of large number of known phases with alkaline-earth or lanthanides instead of alkali metal [9,24,25].

In this work, we have successfully synthesized single crystals of RbCd$_2$Bi$_2$, RbZn$_2$Bi$_2$ and KZn$_2$Bi$_2$ for the first time. As starting point we used recent bismuthides *ATM*Bi self-flux growth method [18] and further optimized it by changing composition of flux. As grown crystals were characterized by combination of powder X-ray diffraction, scanning electron microscopy, energy-dispersive X-ray spectroscopy and single crystal X-ray diffraction. All compounds have gave grown in the same structure type and space group. Our structural investigation yielded with a nice model which describes the differences between cadmium and zinc-based compounds are presented.

## 2. Materials and Methods

All operations with samples and precursors were executed inside of Ar-filled glove box with oxygen and moisture concentration less than 0.1 ppm. Rubidium and potassium 99.5% [GOST 10588-17] were used as is, in the form of glass ampoules. Cadmium or zinc 99.9% [Sigma Aldrich] in the form of drops were used as is, the drops were cut according to the mass requirements. Bismuth 99.98% [GOST 10928-90] in the form of ingot, was milled in agate mortar to small pieces.

Single crystal X-ray analysis was carried out on Bruker D8 QUEST diffractometer (graphite monochromated MoK$\alpha$ radiation, $\omega$-scanning). Crystals were preselected in Ar-filled glove box by using digital microscope Levenhuk DTX 700. For the final selection, directly before the X-ray diffraction analysis, the preselected crystals were taken out of the glove box under layer of dried and degassed liquid paraffin. The intensity data were integrated by the SAINT program and corrected for absorption and decay by SADABS [26]. Structure was solved by dual methods using SHELXT [27] and refined against $F^2$ using SHELXL-2018 [28].

Powder diffraction data for RbZn$_2$Bi$_2$ was taken on Rigaku MiniFlex 600 with Cu K$\alpha$ radiation, equipped with Si monochromator, at room temperature, using sample holder for air-sensitive substances. XRD data was analyzed with Jana2006 software, using Le Bail fit for determination of cell parameters [29].

Microscopic data was collected on JEOL 7001 F scanning electron microscope with an INCA X-act EDS attachment.

Vesta 3 software was used to visualize structures [30].

## 3. Crystal Growth

Crystal growth of $A$TM$_2$Bi$_2$ was achieved with an excess of $A$Bi$_2$ which acts as a self-flux in this case. All the procedures with elements and precursors were executed inside of Ar-filled glovebox. Several crystal growth attempts were performed with various molar ratios for $A$TM$_2$Bi$_2$ – $A$Bi$_2$. Our experiments with the molar ratio $A$TM$_2$Bi$_2$ – $A$Bi$_2$ 3:1 yielded the better growth conditions. For the crystal growth experiments, to achieve $A$TM$_2$Bi$_2$ – $A$Bi$_2$ 3:1 molar ratio, elemental rubidium, or potassium 99.9%, elemental cadmium or zinc 99.9% and bismuth 99.98% were taken in molar ratio 2:3:4 respectively. Masses of reagents were calculated for the total charge of 3.5 g of Bi in every batch, which leads to 4-5.5g total charge. Reagents were loaded in alumina crucible in the order that elements were mentioned above – small balls of alkaline metals first, chunks of transition metals as second layers and bismuth powder was added at the end – to fill free space between pieces of alkaline metal and transition metals as illustrated in Fig. 1(a). Because of the aggressive chemical nature of potassium and rubidium which reacts instantly with bismuth powder if heated up, the mixture should be rammed carefully. Alumina crucible was covered with Nb-foil cap and placed into niobium container, which was welded under 1.05 bar pressure of Ar in arc-welding facility. Further, Nb container was sealed into evacuated silica ampoule with residual pressure $5*10^{-3}$ mbar. Scheme of the assembly is shown on Fig. 1(a).

Also, any oven with homogeneous temperature region can be used and any vertical configuration of assembly is allowed, in our case we used SNOL 10/11-B muffle box furnace. Sealed silica ampoule assembly was inclined on ceramic brick, then the furnace is heated to 900 °C at the rate of 100 degrees per hour, dwelled for 24 hours for better homogenization in liquid state and cooled down to 400 °C at the rate of 2 degrees per hour - to initiate slow nucleation process and crystal growth, as shown in the temperature profile in the Fig. 1(b). Due to layered morphology of the crystals, slow cooling rate is important for optimization of size and the quality of the crystals.

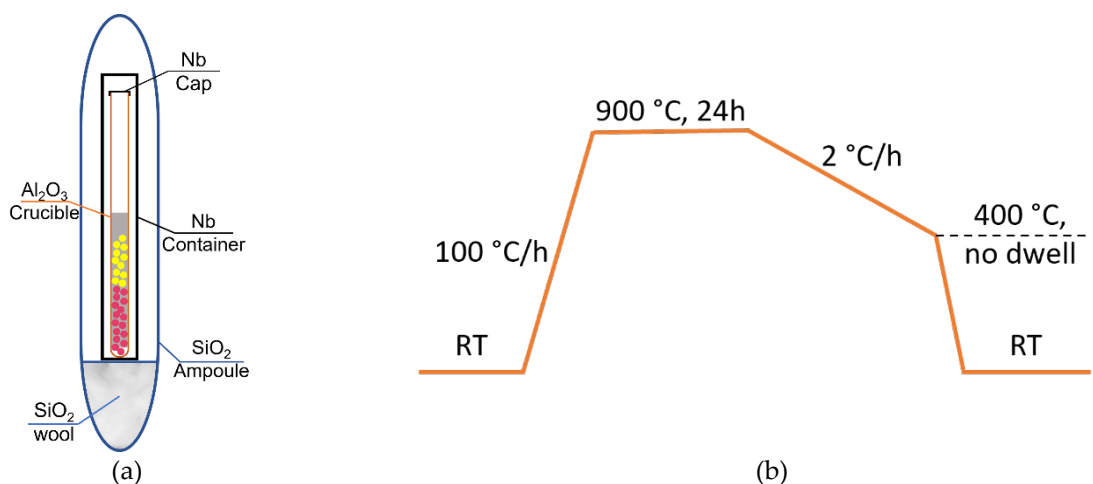

**Figure 1.** Scheme of ampoule for synthesis, where pink circles represent alkaline metal, yellow – transition metal, grey for bismuth powder (a) and thermal profile of synthesis, where RT stands for room temperature (b).

The crucible assembly was opened inside of the glove box, as grown crystals were separated from flux mechanically, using a scalpel and a microscope inside the glovebox, achieving 1.5-2g of crystals. Big blocks of crystals and individual crystals with size up to 4 mm x 5 mm x 0.2 mm were obtained. Crystals can be easily exfoliated with a sharp scalpel or razor blade, perpendicularly to *00l* direction. Exfoliated crystals of RbCd$_2$Bi$_2$ and as-grown crystals of KZn$_2$Bi$_2$ are shown on Fig. 2. Obtained crystals are extremely sensitive to oxygen or moisture and completely degrade on air within seconds.

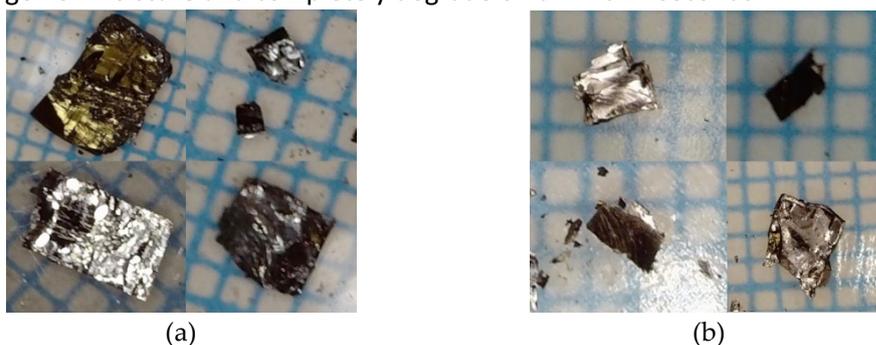

**Figure 2**. As grown crystals after cleaving: RbCd$_2$Bi$_2$ (a) and KZn$_2$Bi$_2$ (b).

## 4. Results and discussion

*4.1 Composition and morphology*

Morphology and composition of RbCd$_2$Bi$_2$ and KZn$_2$Bi$_2$ was studied by using SEM equipped with EDS detector. SEMs images of these as grown crystals are shown on Fig. 3.

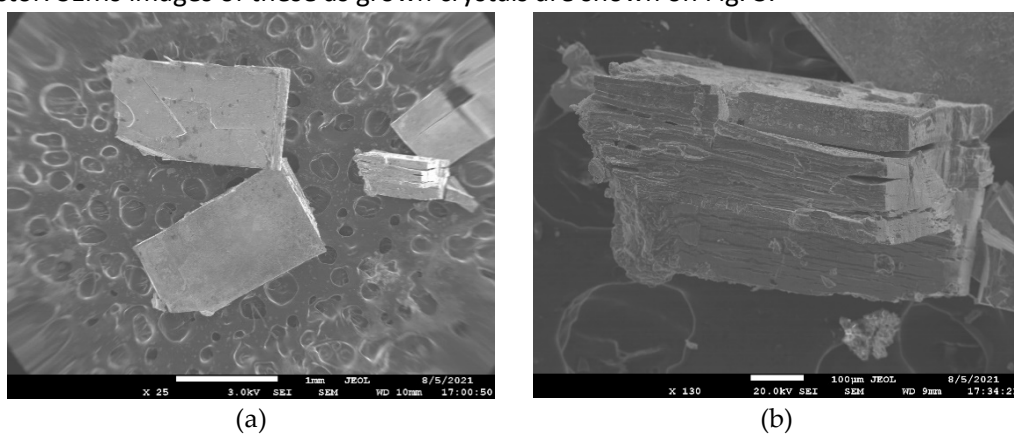

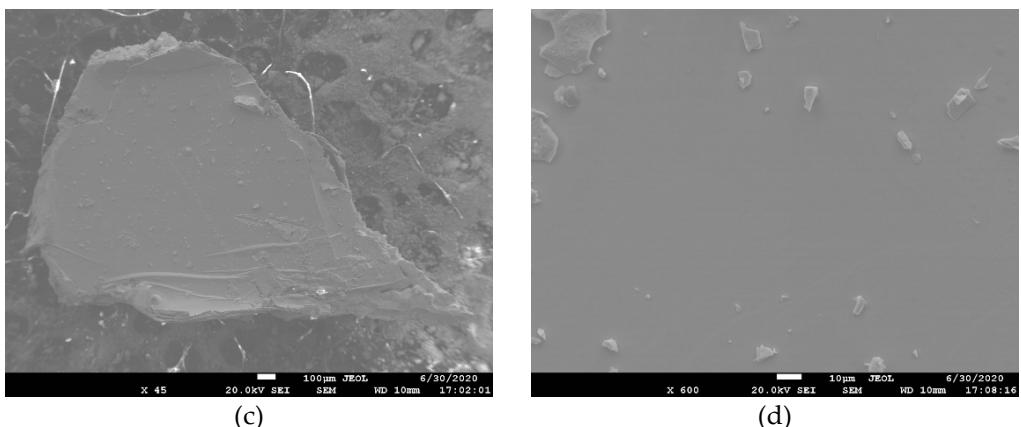

(c)                                 (d)

**Figure 3.** SEM images of as grown $RbCd_2Bi_2$ crystals (a) block of crystals projected perpendicular to the layered structure (b). One large platelet like single crystal (c) and area for EDX analysis (d) of $KZn_2Bi_2$ crystal.

The lamellar crystals with a metallic luster have a regular rectangular shape and were taken as grown. To establish the composition by the EDX, crystals were scanned on several spots, at-least 4-7 spots per crystal. According to the collected data, crystals have the composition $Rb_{1.18(9)}Cd_{1.99(13)}Bi_{1.83(12)}$ and $K_{0.99(2)}Zn_{1.96(3)}Bi_{2.05(3)}$, which, in principle, satisfactorily corresponds to the ideal composition 122, considering the accuracy of the EDX method.

### 4.2 Crystal structure determination

#### 4.2.1 $RbCd_2Bi_2$ crystal structure refinement

Crystallographic data and structure refinement parameters for $RbCd_2Bi_2$ are given in table 1. The coordinates of atoms and other parameters for the structures of $RbCd_2Bi_2$ were deposited with the Cambridge Crystallographic Data Centre (cif file CCDC number 2127201).

**Table 1.** Refinement data for $RbCd_2Bi_2$

| Chemical formula | $RbCd_2Bi_2$ |
|---|---|
| $M_r$ | 760.91 |
| Crystal system, space group | Tetragonal, *I4/mmm* |
| Temperature (K) | 100 |
| $a$, $c$ (Å) | 4.6328(7), 15.217(4) |
| $V$ (Å$^3$) | 326.60(13) |
| Z | 2 |
| Radiation type | Mo K$\alpha$ 0.71073 Å |
| $\mu$ (mm-1) | 67.397 |
| Crystal size, mm | 0.056x0.163x0.195 |
| Data collection | |
| Diffractometer | Bruker Quest D8 |
| Absorption correction | Numerical (SADABS 2016/2) |
| Tmin, Tmax | 0.026, 0.113 |
| No. of measured, independent and observed [I > 2σ(I)] reflections | 2115, 177, 177 |
| Rint | 0.063 |
| Refinement | |
| R, wR, GooF | 0.0242, 0.0559, 1.473 |
| No. of parameters | 9 |
| Largest difference in peak / hole, (e/Å 3 ) | 2.365, -2.097 |
| CCDC | 2127201 |

*4.2.2 KZn$_2$Bi$_2$*

In the case of KZn$_2$Bi$_2$, due to multiple twinning we were unable to refine structure of KZn$_2$Bi$_2$ by single X-ray diffraction. Although all 221 reflections which were harvested from the data set collected by ω-scan were indexed in the same cell (*a*=4.258(2), *c*=15.865(7) Å, *V*=287.6(3)Å$^3$). Furthermore, we were able to solve this structure and refined with surprisingly high residual R1 equal to 19%. Parameter *c* coincided with *00l* powder pattern.

*4.2.3 RbZn$_2$Bi$_2$*

After a few unsuccessful attempts to collect data using X-ray single crystal analysis, due to layered morphology and multiple twinning along *c* axis, in contradiction to KZn$_2$Bi$_2$ we didn`t collect any acceptable data set. In this case, we used powder diffraction x-ray analysis to determine cell parameters of RbZn$_2$Bi$_2$. A block of crystals was grinded with Si powder, which was used to prevent texturing and as internal standard, in agate mortar, obtained powder was charged into thin glass capillary and sealed. Diffraction pattern is shown in Fig. 4. The cell parameters were determined with Le Bail fit, R$_f$ = 0.1245.

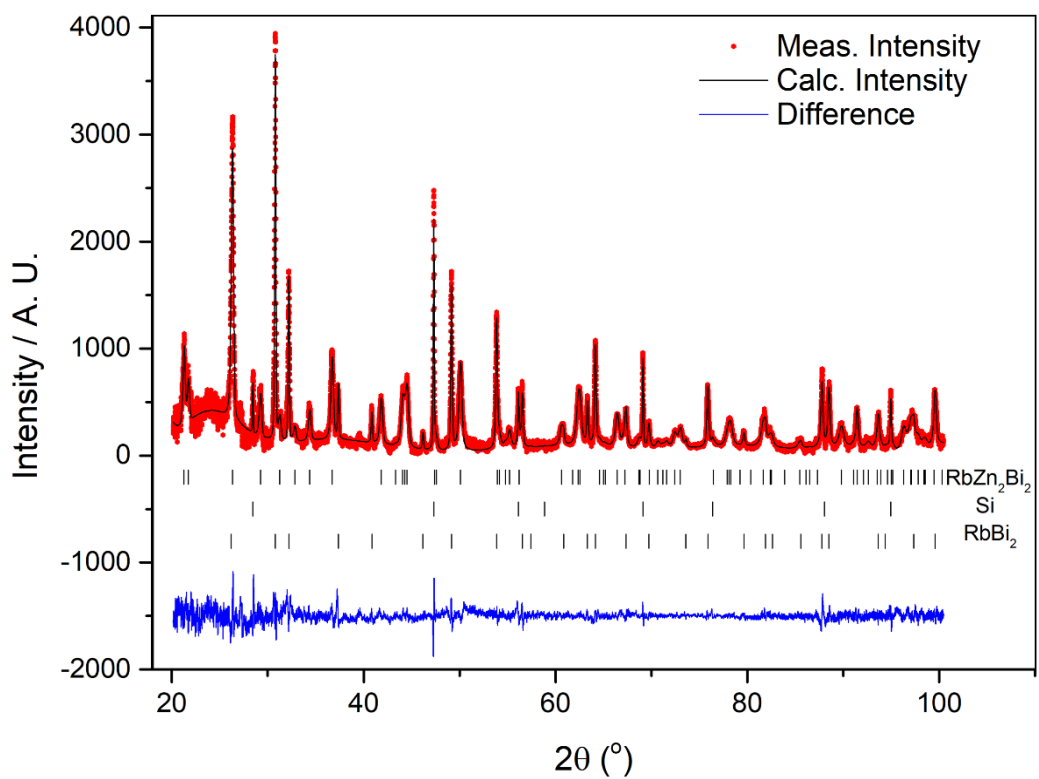

**Figure 4.** Powder X-ray diffraction pattern of RbZn$_2$Bi$_2$. Silicon was taken as internal standard and to prevent main phase from texturing. Le Bail fit was applied to this data.

*4.3 Comparative description of the crystal structures.*

All compounds crystallize in the same structural type ThCr$_2$Si$_2$ (space group *I4/mmm*) and consists of antifluorite-like layers [*TM*Bi] with interlayers of *A* atoms. In this structure type position of transition metal and alkaline metal are fixed, Bi *a* and *b* axis parameters are fixed as well, only *c* axis parameter for Bi atom should be refined. For RbCd$_2$Bi$_2$ and KZn$_2$Bi$_2$ atom positions were determined in structure solution process, in case of RbZn$_2$Bi$_2$ it was calculated by using typical Bi-Zn bond length, which is known for KZn$_2$Bi$_2$ and NaZnBi [18]. Cell parameters, c/a ratio are shown in table 2, bond lengths and valence angles are represented in table 3, atom positions are listened in table 4.

**Table 2.** Cell parameters and c/a ratio.

| Compound | $a$, Å | $c$, Å | $c/a$ |
|---|---|---|---|
| RbCd$_2$Bi$_2$ | 4.6328(7) | 15.217(4) | 3.285 |
| RbZn$_2$Bi$_2$ | 4.3165(2) | 16.3566(6) | 3.789 |
| KZn$_2$Bi$_2$ | 4.258(2) | 15.865(7) | 3.726 |

**Table 3.** Angles and bond distances for $ATM_2Bi_2$

| | RbCd$_2$Bi$_2$ | RbZn$_2$Bi$_2$ | KZn$_2$Bi$_2$ |
|---|---|---|---|
| 4xBi-TM-Bi | 113.489(12)° | 114.51(17)° | 115.03(16)° |
| 2xBi-TM-Bi | 101.70(3)° | 99.8(4)° | 98.8(3)° |
| Bi-$TM$, Å | 2.9869(6) | 2.821(7) | 2.801(6) |
| Bi-$A$, Å | 3.7963(6) | 3.805(6) | 3.698(5) |
| Bi--Bi, Å | 3.8371(14) | 4.54(2) | 4.30(2) |

**Table 4.** Atom positions for $ATM_2Bi_2$

| | | x/$a$ | y/$b$ | z/$c$ |
|---|---|---|---|---|
| $A$ | | 0.5 | 0.5 | 0 |
| $TM$ | | 0.5 | 0 | 0.25 |
| Bi | RbCd$_2$Bi$_2$ | 0.5 | 0.5 | 0.6260(1) |
| | RbZn$_2$Bi$_2$ | | | 0.6389(5) |
| | KZn$_2$Bi$_2$ | | | 0.6354(6) |

In antifluorite-like layers, the *TM* atoms form a planar square network, and the Bi atoms are arranged in staggered rows above and under the centers of the [*TM*4] squares due to which each *TM* atom has a tetrahedral environment of four Bi atoms. RbCd$_2$Bi$_2$ structure is represented on Fig. 5a.

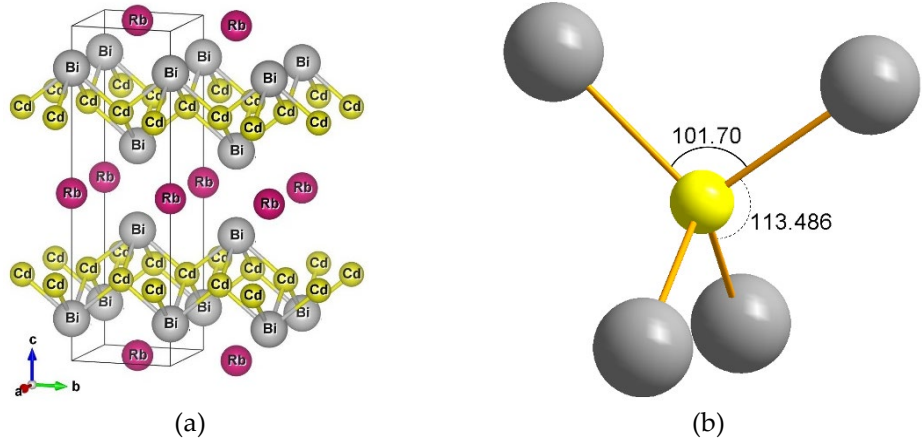

(a)          (b)

**Figure 5.** a) General view of RbCd$_2$Bi$_2$. b) Tetrahedron [CdBi$_4$], angles show deformation of tetrahedron in RbCd$_2$Bi$_2$ structure as a result of some stretching along the c axis

*TM*-Bi bond length increasing correlates with increasing of effective radii of *TM* for Cd and Zn containing compounds. The tetrahedral polyhedron [*TM*Bi4] is somewhat extended along the inversion C4i axis passing parallel to the c axis through two opposite edges of the tetrahedron Fig. 5b. This is indicated by the values of two Bi*TM*Bi angles leaned on these edges. The Rb/K atoms are arranged between the [*TM*Bi] layers to form the coordination polyhedron [$A$Bi$_8$] as a tetragonal prism. The bismuth atom is placed inside the tetragonal antiprism [Bi$A_4TM_4$]. The bottom base of the prism is formed by four A atoms from one layer, and the upper base are formed by four *TM* atoms. The [Bi$A_4TM_4$] polyhedron is oriented along the *c* axis, the C4 axis of rotation passes through the middle of the antiprism bases.

Bi-Bi distance dramatically decreases from $A$Zn$_2$Bi$_2$ to RbCd$_2$Bi$_2$, that explains difference in crystal morphology and twinning along *c* axis.

An interesting problem with formal charges in these compounds, which bounds with inaccessible for Cd and Zn oxidation state TM$^{+3}$. For known structures such as CsMn$_2$P$_2$ it was found, that transition metal contains two different oxidation states - $Mn^{+2}$ and $Mn^{+3}$, which was confirmed by magnetic susceptibility measurements [23]. The same phenomena was found for $RbFe_2As_2$ [2]. For our compounds, with the $d^{10}$ type of transition metals this way of describing the formal charges can`t be applied.

Transition from zinc to cadmium leads to growth on *a* and *b* directions and to squeeze on *c* direction, which can be associated with difference in electronegativity of cadmium and zinc – 1.46 and 1.66 respectively [31]. Compared to electronegativity of bismuth which equals to 1.67, we can assume, that Bi-Cd bonds are slightly polar, but in Bi-Zn layer all bonds are covalent. Same process of interacting layers can be found in transition from $SrCo_2As_2$ to $SrNi_2As_2$ [32], when uncollapsed (antifluorite-like layers does not interact) tetragonal structure transforms into collapsed (antifluorite-like layers interact) with substitution of *TM*. Difference in structure between these types is shown in Fig. 6.

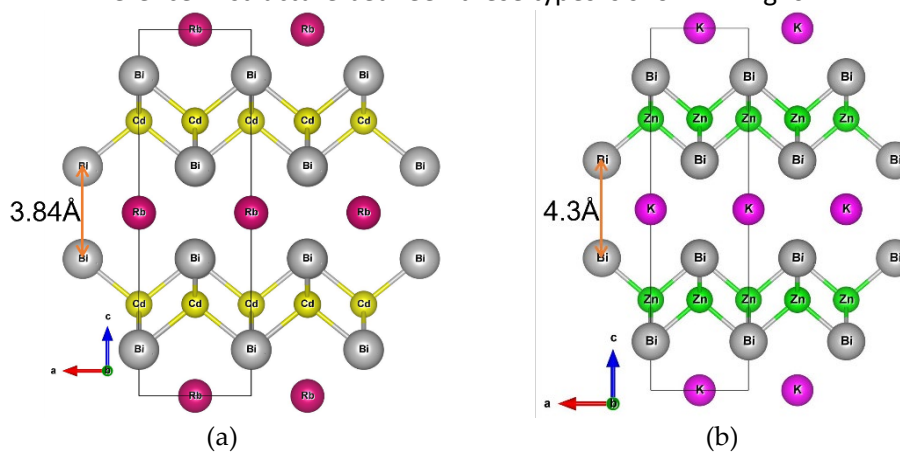

(a)          (b)

**Figure 6.** (a) View towards *b-axis* of $RbCd_2Bi_2$ (b) $KZn_2Bi_2$ towards *b-axis*. On figures 7(a, b) Bi-Bi distance marked with orange dual-arrow.

## 5. Conclusions

We report optimized crystal growth conditions for the ternary compounds $RbCd_2Bi_2$, $KZn_2Bi_2$, $RbZn_2Bi_2$ via self-flux method. A molar ratio of reagents and a temperature profile were optimized to get crystals up-to 4-5 mm in size. As grown crystals can be easily separated from flux. The as-grown crystals have plate-like shape, layered morphology and tetragonal habitus, their composition was confirmed by combination of X-ray diffraction methods and EDX spectroscopy. $RbCd_2Bi_2$ structure was solved and refined with single-crystal x-ray diffraction. Our refined model for this compound confirms that it crystallizes in *I4/mmm* space group. $AZn_2Bi_2$ are isostructural to $RbCd_2Bi_2$ according to single-crystal x-ray diffraction for $KZn_2Bi_2$ and x-ray powder pattern for $RbZn_2Bi_2$.

Short Bi-Bi distance in $RbCd_2Bi_2$ and mixed oxidation states for cadmium atoms are promising to yield interesting electronic and magnetic properties, and for better understanding of Bi-*TM* bonding in these compounds. Furthermore, electron-deficiency in $RbCd_2Bi_2$ can lead to substitution of Rb to $AE^{+2}$ and Cd to $TM^{+3}$ in search for materials with intriguing properties.

Single crystals of $ATM_2Bi_2$ that were grown in this work are currently under investigation for physical properties such as magnetic and electronic structure properties, which can lead to the understating of their ground states of these new materials.

**Funding:** A.I.S., D.V.E., B.B., S.A. and I.V.M thank DFG (#405940956) and RSF (#19-43-04129) for financial support in the frame of the joint DFG-RSF project "Weyl and Dirac semimetals and beyond –prediction, synthesis and characterization of new semimetals", S.A. acknowledges DFG through Grant No: AS 523/4-1.

**Acknowledgments:** Electronic images and powder patterns were obtained using the facilities of the International Laboratory of High Magnetic Fields and Low Temperatures and the Shared Research

Facilities Center at P N Lebedev Physical Institute. The authors acknowledge support from Lomonosov Moscow State University Program Development for providing access to single x-ray diffraction data.# 6. References

1. Hosono, H. Two Classes of Superconductors Discovered in Our Material Research: Iron-Based High Temperature Superconductor and Electride Superconductor. Physica C: Superconductivity 2009, 469, 314–325, doi:10.1016/j.physc.2009.03.014.
2. Moroni, M.; Prando, G.; Aswartham, S.; Morozov, I.; Bukowski, Z.; Büchner, B.; Grafe, H.J.; Carretta, P. Charge and Nematic Orders in AFe2As2 (A = Rb,Cs) Superconductors. Phys. Rev. B 2019, 99, 235147, doi:10.1103/PhysRevB.99.235147.
3. Pitcher, M.J.; Parker, D.R.; Adamson, P.; Herkelrath, S.J.C.; Boothroyd, A.T.; Ibberson, R.M.; Brunelli, M.; Clarke, S.J. Structure and Superconductivity of LiFeAs. Chem. Commun. 2008, 5918, doi:10.1039/b813153h.
4. Alireza, P.L.; Ko, Y.T.C.; Gillett, J.; Petrone, C.M.; Cole, J.M.; Lonzarich, G.G.; Sebastian, S.E. Superconductivity up to 29 K in SrFe2As2 and BaFe2As2 at High Pressures. J. Phys.: Condens. Matter 2009, 21, 012208, doi:10.1088/0953-8984/21/1/012208.
5. Kihou, K.; Saito, T.; Ishida, S.; Nakajima, M.; Tomioka, Y.; Fukazawa, H.; Kohori, Y.; Ito, T.; Uchida, S.; Iyo, A.; et al. Single Crystal Growth and Characterization of the Iron-Based Superconductor KFe2As2 Synthesized by KAs Flux Method. J. Phys. Soc. Jpn. 2010, 79, 124713, doi:10.1143/JPSJ.79.124713.
6. Gvozdetskyi, V.; Owens-Baird, B.; Hong, S.; Zaikina, J. Thermal Stability and Thermoelectric Properties of NaZnSb. Materials 2018, 12, 48, doi:10.3390/ma12010048.
7. Guo, K.; Cao, Q.; Zhao, J. Zintl Phase Compounds AM2Sb2 (A=Ca, Sr, Ba, Eu, Yb; M=Zn, Cd) and Their Substitution Variants: A Class of Potential Thermoelectric Materials. Journal of Rare Earths 2013, 31, 1029–1038, doi:10.1016/S1002-0721(12)60398-6.
8. Yang, J.; Wegner, A.; Brown, C.M.; Louca, D. Defect-Driven Extreme Magnetoresistance in an I-Mn-V Semiconductor. Appl. Phys. Lett. 2018, 113, 122105, doi:10.1063/1.5040364.
9. Calder, S.; Saparov, B.; Cao, H.B.; Niedziela, J.L.; Lumsden, M.D.; Sefat, A.S.; Christianson, A.D. Magnetic Structure and Spin Excitations in BaMn 2 Bi 2. Phys. Rev. B 2014, 89, 064417, doi:10.1103/PhysRevB.89.064417.
10. Li, L.; Wang, K.; Graf, D.; Wang, L.; Wang, A.; Petrovic, C. Electron-Hole Asymmetry, Dirac Fermions, and Quantum Magnetoresistance in BaMnBi2. Phys. Rev. B 2016, 93, 115141, doi:10.1103/PhysRevB.93.115141.
11. Thirupathaiah, S.; Efremov, D.; Kushnirenko, Y.; Haubold, E.; Kim, T.K.; Pienning, B.R.; Morozov, I.; Aswartham, S.; Büchner, B.; Borisenko, S.V. Absence of Dirac Fermions in Layered BaZnBi2. Phys. Rev. Materials 2019, 3, 024202, doi:10.1103/PhysRevMaterials.3.024202.
12. Ishizaka, K.; Bahramy, M.S.; Murakawa, H.; Sakano, M.; Shimojima, T.; Sonobe, T.; Koizumi, K.; Shin, S.; Miyahara, H.; Kimura, A.; et al. Giant Rashba-Type Spin Splitting in Bulk BiTeI. Nature Mater 2011, 10, 521–526, doi:10.1038/nmat3051.
13. Qu, D.-X.; Hor, Y.S.; Xiong, J.; Cava, R.J.; Ong, N.P. Quantum Oscillations and Hall Anomaly of Surface States in the Topological Insulator Bi2Te3. Science 2010, 329, 821–824, doi:10.1126/science.1189792.
14. Kushwaha, S.K.; Krizan, J.W.; Feldman, B.E.; Gyenis, A.; Randeria, M.T.; Xiong, J.; Xu, S.-Y.; Alidoust, N.; Belopolski, I.; Liang, T.; et al. Bulk Crystal Growth and Electronic Characterization of the 3D Dirac Semimetal Na3Bi. APL Materials 2015, 3, 041504, doi:10.1063/1.4908158.
15. Borisenko, S.; Gibson, Q.; Evtushinsky, D.; Zabolotnyy, V.; Büchner, B.; Cava, R.J. Experimental Realization of a Three-Dimensional Dirac Semimetal. Phys. Rev. Lett. 2014, 113, 027603, doi:10.1103/PhysRevLett.113.027603.
16. Thirupathaiah, S.; Morozov, I.; Kushnirenko, Y.; Fedorov, A.V.; Haubold, E.; Kim, T.K.; Shipunov, G.; Maksutova, A.; Kataeva, O.; Aswartham, S.; et al. Spectroscopic Evidence of Topological Phase Transition in the Three-Dimensional Dirac Semimetal Cd3(As1−xPx)2. Phys. Rev. B 2018, 98, 085145, doi:10.1103/PhysRevB.98.085145.
17. Lu, H.; Zhang, X.; Bian, Y.; Jia, S. Topological Phase Transition in Single Crystals of (Cd1−xZnx)3As2. Sci Rep 2017, 7, 3148, doi:10.1038/s41598-017-03559-2.
9